\documentclass[conference, 10pt]{IEEEtran}
\IEEEoverridecommandlockouts
\usepackage[T1]{fontenc}
\usepackage[utf8]{inputenc}
\usepackage[english]{babel}
\usepackage{balance}
\usepackage{cite}
\usepackage{amsmath, mathtools}
\usepackage{amsfonts,amsthm,bm}
\usepackage{amssymb}
\usepackage{comment}
\usepackage{graphicx}
\usepackage{standalone}
\usepackage{dsfont}
\usepackage{mathtools}
\usepackage{color, colortbl}
\usepackage{soul}
\usepackage[normalem]{ulem}
\usepackage[acronym,shortcuts]{glossaries}
\usepackage{arydshln}
\usepackage{bbm}
\usepackage{svg} 
\usepackage{algorithm,algorithmic}
\usepackage{hyperref}
\usepackage{balance}
\usepackage{adjustbox}
\usepackage[inline]{enumitem}

\usepackage{tikz}
\usepackage{pgfplots}
\pgfplotsset{compat=newest}
\pgfplotsset{plot coordinates/math parser=false}
\newlength\fheight
\newlength\fwidth
\usetikzlibrary{plotmarks,patterns,decorations.pathreplacing,backgrounds,calc,arrows,arrows.meta,spy,matrix}
\usepgfplotslibrary{patchplots,groupplots}
\usepackage{tikzscale}

\usetikzlibrary {patterns.meta}

\DeclareMathOperator*{\argmax}{\arg\!\max}
\DeclareMathOperator*{\argmin}{\arg\!\min}

\DeclareMathOperator*{\diag}{diag}

\newacronym{3gpp}{3GPP}{3rd Generation Partnership Project}
\newacronym{5g}{5G}{5th generation}
\newacronym{af}{AF}{amplify-and-forward}
\newacronym{cwc}{CWC}{capacity-weighted clustering}
\newacronym{ecdf}{ECDF}{empirical cumulative distribution function}
\newacronym{fov}{FoV}{field-of-view}
\newacronym{gnb}{gNB}{next generation node base}
\newacronym{hc}{HC}{hierarchical clustering}
\newacronym{km}{KM}{K-means}
\newacronym{kmed}{KMed}{K-medoids}
\newacronym{iab}{IAB}{integrated access and backhaul}
\newacronym{irs}{IRS}{intelligent reflecting surface}
\newacronym{isd}{ISD}{inter-site distance}
\newacronym{los}{LoS}{line-of-sight}
\newacronym{minlp}{MINLP}{mixed integer nonlinear programming}
\newacronym{mimo}{MIMO}{multiple input-multiple output}
\newacronym{mmwave}{mmWave}{millimeter wave}
\newacronym{nlos}{NLoS}{non-line-of-sight}
\newacronym{oscwc}{OS-CWC}{one shot \ac{cwc}}
\newacronym{pam}{PAM}{partition around medoids}
\newacronym{rb}{RB}{resource block}
\newacronym{snr}{SNR}{signal-to-noise-ratio}
\newacronym{sinr}{SINR}{signal-to-interference-plus-noise-ratio}
\newacronym{svd}{SVD}{singular value decomposition}
\newacronym{tdma}{TDMA}{time division multiple access}
\newacronym{tti}{TTI}{transmission time interval}
\newacronym{thz}{THz}{terahertz}
\newacronym{ue}{UE}{user equipment}
\newacronym{ula}{ULA}{uniform linear array}
\newacronym{uma}{UMa}{urban macro-cell}
\newacronym{umi}{UMi}{urban micro-cell}
\newacronym{upa}{UPA}{uniform planar array}

\title{Downlink \acs{tdma} Scheduling for \acs{irs}-aided Communications with Block-Static Constraints}
\author{ Alberto Rech$^*$,  Matteo Pagin$^*$, Stefano Tomasin$^*$, Federico Moretto$^*$,  \\ Leonardo Badia$^*$, Marco Giordani$^*$, Jonathan Gambini$^\dagger$, and Michele Zorzi$^*$\\
\small $^*$Department of Information Engineering, University of Padova, Italy. \\
$^\dagger$Milan Research Center, HUAWEI, Italy.}

\begin{document}
\maketitle

\begin{abstract}
\Acp{irs} are being studied as possible low-cost energy-efficient alternatives to active relays, with the goal of solving the coverage issues of \ac{mmwave} and \ac{thz} network deployments. In the literature, these surfaces are often studied by idealizing their characteristics. Notably, it is often assumed that \acp{irs} can tune with arbitrary frequency the phase-shifts induced by their elements, thanks to a wire-like control channel to the \ac{gnb}. Instead, in this work we investigate an \ac{irs}-aided \ac{tdma} cellular network, where the reconfiguration of the \ac{irs} may entail an energy or communication cost, and we aim at limiting the number of reconfigurations over time. We develop a clustering-based heuristic scheduling, which optimizes the system sum-rate subject to a given number of reconfigurations within the TDMA frame. To such end, we first cluster \acp{ue} with a similar optimal \ac{irs} configuration. Then, we compute an overall \ac{irs} cluster configuration, which can be thus kept constant while scheduling the whole \acp{ue} cluster.
Numerical results show that our approach is effective in supporting \acp{irs}-aided systems with practical constraints, achieving up to $85$\% of the throughput obtained by an ideal deployment, while providing a $50$\% reduction in the number of \ac{irs} reconfigurations.

\end{abstract}


\glsresetall
\IEEEpeerreviewmaketitle

\section{Introduction}\label{sec:introduction}

The ever-increasing growth of mobile traffic has called both academia and industry to identify and develop solutions for extending the radio spectrum beyond the crowded sub-6 GHz bands. As a result of these efforts, the latest iteration of the cellular standard, i.e., 5G NR, has introduced the support for communication in the \ac{mmwave} bands~\cite{38104}. Moreover, the use of \ac{thz} frequencies is being investigated as a possible key technology enabler for 6G networks as well~\cite{polese2020toward}.

However, \ac{mmwave} and beyond frequencies exhibit challenging propagation conditions, mainly due to the severe path loss and susceptibility to blockages~\cite{rangan2017potentials}.
To mitigate these limitations, a possible solution is to densify the network, i.e., to reduce the cell radius of 5G and beyond base stations with respect to sub-6 GHz ones. Unfortunately, this approach is proving to be unfeasible for network operators, since trenching and deploying the necessary fiber backhaul links usually represents a financial and logistical hurdle~\cite{lopez2015towards}. This issue is exacerbated in remote areas, where the limited access to electrical power and the lower \ac{ue} density limit even further the feasibility of deploying dense networks from a business standpoint~\cite{chaoub20216g}.


In light of this, \acp{irs} are being investigated as possible solutions to overcome the harsh propagation conditions exhibited by \ac{mmwave} and \ac{thz} bands in a cost- and energy-efficient manner~\cite{flamini2022towards}. 
Specifically, \acp{irs} are network entities whose radiating elements can passively tune the phase-shift of impinging signals. Therefore, they can be used to beamform the reflected signal towards a virtually arbitrary destination (i.e., the receiver), hence improving the signal quality without an active amplification~\cite{bjornson2019intelligent}. 

Despite the substantial research hype, it must be noted that most studies that consider \acp{irs} as promising solutions for limiting \ac{mmwave} and \ac{thz} coverage holes rely on assumptions which are unlikely to be satisfied in actual real-world deployments. Specifically, a significant body of literature studies \acp{irs} under the premise of the presence of an ideal control channel among the latter and the base station~\cite{8910627, 9115725, 8811733, 9810370}. Instead, actual deployments will likely feature a wireless \ac{irs} control channel, possibly implemented with low-cost and low-power technologies~\cite{liu2022path, liaskos2018realizing}. 
In turn, this will introduce constraints on the reconfiguration period of the \ac{irs}, which needs to be synchronized with the base station in order to beamform the signal towards the \ac{ue} served during the specific \ac{tti}~\cite{flamini2022towards}. 

Additionally, early \ac{irs} control circuitry prototypes, which indeed have a low power consumption (i.e., in the order of hundreds of mW), exhibit non-negligible phase-shifts reconfiguration time~\cite{rossanese2022designing}. Thus, a constraint on the re-configuration period should be accounted for to ensure system synchronization. In this regard, it is of interest to 
\begin{enumerate*}[label=(\textit{\roman*})]
  \item investigate the magnitude of the performance degradation experienced by \ac{irs}-aided systems when considering such practical constraints and
  \item design algorithms that aim at mitigating these limitations. 
\end{enumerate*}

The whole problem of assigning resources with multifaceted parameters in a potentially large network is not new to cellular scheduling, in particular it can be found to coordinated multipoint (CoMP) or similar multidimensional allocations, where it can be solved by assigning repeated patterns so as to save complexity \cite{guidolin2013statistical,marotta2019network}, and at the same time identifying allocation clusters in a distributed fashion \cite{bassoy2016load,guidolin2014distributed}.

In a sense, we can think of exploiting these ideas in the entirely different context of designing a low-cost control for an IRS. In more detail, we consider a \ac{tdma}-scheduled communication system in which an \ac{irs}, shared among multiple \acp{ue} to 
improve their performance in terms of end-to-end \ac{snr}, can be reconfigurated a limited number of times per radio time frame. To this end, we propose a time division scheduling policy, based on clustering algorithms, which aims to provide throughput enhancements in \ac{irs}-aided network deployments with practical constraints. In particular, we present the \ac{cwc} technique, which is shown to be very effective in guaranteeing high system sum-rate performance despite the aforementioned constraints.

The rest of the paper is organized as follows. In Section~\ref{sec:system_model}, we introduce the system model and we present the sum-rate optimization problem. In Section~\ref{sec:optimization}, we provide a heuristic solution aiming at maximizing the system sum-rate. In Section~\ref{sec:numerical_results}, we discuss the numerical results and compare different scheduling solutions. Finally, Section~\ref{sec:conclusions} draws the main conclusions.

{\it Notation:} Scalars are denoted by italic letters, vectors and matrices  by boldface lowercase and uppercase letters, respectively, and sets are denoted by calligraphic uppercase letters. $\diag(\bm{a})$ indicates a square diagonal matrix with the elements of $\bm{a}$ on the principal diagonal. $\bm{A}^H$ denotes the conjugate transpose of matrix $\bm{A}$.
Finally, $[\bm{A}]_{ij}$ denotes the scalar value in the $i$-th row and $j$-th column of matrix $\bm{A}$. 

\section{System Model}\label{sec:system_model}

\begin{figure}
    \centering
    \includegraphics[width=\linewidth]{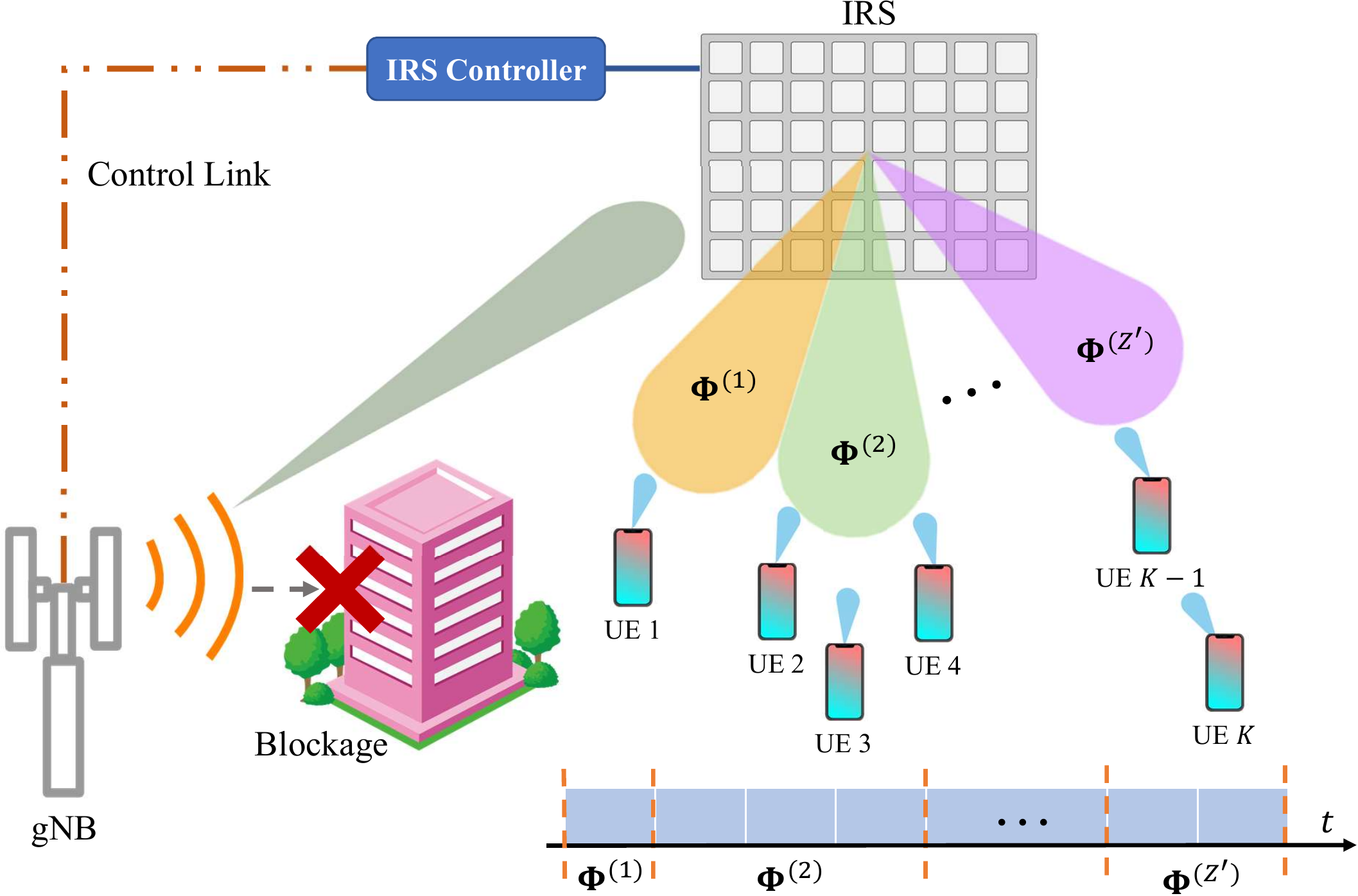}
    \caption{\ac{tdma} scheduling for \ac{irs}-assisted multi-\ac{ue} communication.}
    \label{fig:system_model}
\end{figure}

We consider downlink data transmission for the multi-\ac{ue} \ac{mimo} communication system shown in Fig.~\ref{fig:system_model}, wherein the transmission from the \ac{gnb} to the $K$ \acp{ue} is assisted by an \ac{irs}.

The \ac{gnb} and the \acp{ue} are equipped with $N_{\rm g}$ and $N_{\rm k}$ antennas, respectively. We assume that the system operates either in the \ac{mmwave} or the \ac{thz} bands and that the direct link between the \ac{gnb} and the \acp{ue} is unavailable due to a deep blockage. As a consequence, the \ac{gnb} transmits signals to the \acp{ue} by exploiting the virtual \ac{los} link offered by the \ac{irs}.
Time is divided into frames, each split into $K$ slots, and each \ac{ue} is served exactly once in a frame. 

\paragraph*{\ac{irs} Model}
The $N_{\rm I}$ elements of the \ac{irs} act each as an omnidirectional antenna unit that reflects the impinging electromagnetic field, by introducing a tunable phase shift on the baseband-equivalent signal model. We denote with $\phi_n=e^{j\theta_n}$ the reflection coefficient of the $n$-th \ac{irs} element, where $\theta_n \in [-\pi,\pi)$ is the induced phase shift. Since recent works argue that continuous-phase shifts are hardly implementable in practice \cite{Tan2018}, we also consider the case of quantized configurations, in which phase shifts are chosen from a discrete set $\mathcal{P}_{\theta} = \left\{0, \frac{2\pi}{2^b},\ldots,\frac{2\pi(2^b-1)}{2^b} \right\}$, being $b>0$ the number of bits employed to control the quantized phase shifts.

We denote with $\bm{H}(f) \in \mathbb{C}^{N_{\rm I} \times N_{\rm g}}$ the \ac{gnb}-\ac{irs} channel matrix and with $\bm{G}_k(f) \in \mathbb{C}^{N_{\rm U} \times N_{\rm I}}$ the channel matrix of the link between the \ac{irs} and \ac{ue} $k$.
We consider single-stream transmissions, with $\bm{w}_{{\rm g}_k} \in \mathbb{C}^{N_{\rm g}\times 1}$ and $\bm{w}_{{\rm U}_k}\in \mathbb{C}^{N_{\rm U}\times 1}$ defined as the beamforming vectors at the \ac{gnb} and the \ac{ue} $k$. 
Let $x_k$ be the single-stream signal transmitted by the \ac{gnb} to \ac{ue} $k$, the received signal can be then expressed as
\begin{equation}
    z_k = \bm{w}_{{\rm U}_k}^{T} \bm{G}_k \bm{\Phi} \bm{H}  \bm{w}_{{\rm g}_k}x_k + \bm{w}_{{\rm U}_k}^{T} \bm{n}_{k}.
\end{equation}
where $\bm{n}_k\in \mathbb{C}^{N_{\rm U}\times 1}$ represents the circularly symmetric complex Gaussian noise vector with zero-mean and variance $\sigma^2_{\rm n}$ and $\bm{\Phi} \in \mathbb{C}^{N_{\rm I}\times N_{\rm I}}$ is the {\em \ac{irs} configuration}, i.e., a diagonal matrix defined as $\bm{\Phi} = \diag(\phi_1,\ldots,\phi_{N_{\rm I}})$. Note that specific \ac{irs} configurations can be adopted for different \acp{ue}. Accordingly, in the following $\bm{\Phi}(k)$ denotes the configuration adopted when \ac{ue} $k$ is served.

The \ac{snr} at \ac{ue} $k$, with \ac{irs} configuration $\bm{\Phi}(k)$, is given by
\begin{equation}\label{snr_k}
    \Gamma_k(\bm{\Phi}(k)) = \frac{|\bm{w}_{{\rm U}_k}^{T} \bm{G}_k \bm{\Phi}(k) \bm{H}  \bm{w}_{{\rm g}_k} |^2\sigma_{\bm{x}}^2}{|\bm{w}_{{\rm U}_k}|^2\sigma_{\bm{n}}^2},
\end{equation}
where $\sigma_{\bm{x}}^2$ is the signal power.
In general, different \ac{irs} configurations should be adopted for each \ac{ue} to maximize its own\ac{snr}, based on its position in the cell and the channel conditions. 
The goal of this paper, however, is to limit the IRS reconfigurations as discussed in Section~\ref{sec:introduction}. Heuristic algorithms able to maximize \acp{ue} performance while complying with this requirement will be presented in Section~\ref{sec:optimization}.

\subsection{Sum-rate Optimization Problem}

The number of \ac{irs} reconfigurations may be limited, with the goal of accounting for practical limitations that might arise in realistic deployments.

\begin{enumerate*}[label=\roman*]
\item reducing the power consumption of the control circuitry and 
\item taking into account for the limitations incurred by a realistic control channel to the \ac{gnb}. 
\end{enumerate*}
On the downside, achieving this goal usually leads to \ac{snr} degradation as sub-optimal \ac{irs} configurations might be adopted to serve some \acp{ue}. To mitigate this effect, we formulate a constrained optimization problem on the average system sum-rate.
In the following, the constraint on the maximum number of reconfigurations within a time frame will be referred to as the \emph{block-static constraint}. For the latter, we assume the following:
\begin{enumerate}[label=\arabic*.]
    \item \textit{at most} $Z$ \ac{irs} reconfigurations can occur within a time frame;
    \item the \ac{gnb} serves the $K$ \acp{ue} by partitioning them into $Z$ disjoint subsets $\mathcal{U}_1,\ldots,\mathcal{U}_{Z}$, with $Z\leq K$;
    \item for each \acp{ue} subset $\mathcal{U}_z$, the same \ac{irs} configuration $\bm{\Phi}^{(z)}$ is kept, i.e., $\bm{\Phi}(k) = \bm{\Phi}^{(z)}, \forall k \in \mathcal{U}_z, \forall \, 1\leq z \leq Z$.
\end{enumerate}
Let $\mathcal{C} = \{\bm{\Phi}^{(1)}, \bm{\Phi}^{(2)}, \ldots, \bm{\Phi}^{(Z)}\}$ be the set of \ac{irs} configurations corresponding to subsets $\mathcal{U}_1,\ldots,\mathcal{U}_{Z}$. The average system sum-rate within a time frame is defined as
\begin{equation}
    R(\mathcal{U}_1,\ldots,\mathcal{U}_{Z},\mathcal{C}) = \sum_{z=1}^{Z} \sum_{k \in \mathcal{U}_z}\log_2\left(1+\Gamma_k(\bm{\Phi}^{(z)}\right) ,
\end{equation} 
where $\Gamma_k(\bm{\Phi}^{(z)})$ is the \ac{snr} experienced by the $k$-th \ac{ue} when using the \ac{irs} configuration $\bm{\Phi}^{(z)}$.

The optimization problem is thus formulated as
\begin{subequations}\label{optproblem}
    \begin{equation}
        \argmax_{\substack{\mathcal{U}_1,\ldots,\mathcal{U}_{Z}, \mathcal{C}}} R(\mathcal{U}_1,\ldots,\mathcal{U}_{Z},\mathcal{C}),
        \\
    \end{equation}
    \vspace{-10pt}
    \begin{alignat}{2}
     & \text{s.t.}\; & \theta_{n, z} \in [-\pi, \pi),\label{const_phaseshifters}
    \end{alignat}
\end{subequations}
where $\theta_{n, z} = \angle[\bm{\Phi}^{(z)}]_{n,n}$, for $n =1,\ldots,N_{\rm I}$.

\section{Constrained Sum-rate Optimization}\label{sec:optimization}

In this section, we provide a heuristic solution to \eqref{optproblem}. Specifically, we first present two clustering-based approaches to identify and group \acp{ue} with a similar optimal \ac{irs} configuration. Then, we solve the scheduling problem on the identified clusters with a TDMA approach.
We compute the \acp{ue} clusters by first estimating the optimal \textit{individual \ac{irs} configuration}, i.e., the configurations leading to the maximum rate when considering only one of the \acp{ue} in Sec. \ref{sec:inv_opt}. 
These configurations would solve \eqref{optproblem} for 
$Z = K$, as in this case all \acp{ue} are served in a \ac{tdma} fashion and with their optimal \ac{irs} configuration, denoted as $\bm{\Phi}^*(k)$. 
The phase coefficients of the optimal \ac{irs} configuration matrices are then chosen as the initial points of a procedure leveraging arbitrary clustering algorithms in the $N_{\rm I}$-dimensional space, as explained in Sec. \ref{sec:inv_opt}. Finally, we propose ad hoc clustering techniques in Sec. \ref{sec:clust_cwc}.

\subsection{Individual Optimal IRS Configurations}
\label{sec:inv_opt}
In \ac{mimo} systems, both \ac{gnb} and \acp{ue} adopt properly tuned beamformers to steer the signal towards the spatial direction providing the highest channel gain \cite{Tomasin21}. For the optimization of the \ac{irs} configuration of each individual \ac{ue}, we adopt a procedure similar to the one presented in \cite{Qian22joint}, focusing on the single-stream transmissions. 
The optimal beamforming vectors  $\bm{w}_{{\rm U}_k}$ and $\bm{w}_{{\rm g}_k}$ coincide with the singular vectors corresponding to the highest singular value of the wireless channel matrix. In principle, one could thus estimate  $\bm{w}_{{\rm U}_k}$ and $\bm{w}_{{\rm g}_k}$ by applying singular value decomposition on the overall cascade channel matrix
\begin{equation}\label{eq:svd}
    \bm{G}_k \bm{\Phi}(k) \bm{H} = \bm{U}\bm{\Sigma}\bm{V}^H,
\end{equation}
and obtaining the right and left singular vectors of $\bm{G}_k \bm{\Phi}(k) \bm{H}$ as the columns of $\bm{V}$ and $\bm{U}$ and the corresponding singular values as the diagonal entries of $\bm{\Sigma}$.
Notice, though, that the cascade channel itself depends on the specific \ac{irs} configuration $\bm{\Phi}(k)$, which in our formulation represents one of the optimization variables.
Indeed, for fixed $\bm{w}_{{\rm U}_k}$ and $\bm{w}_{{\rm g}_k}$ we can solve
\begin{equation}\label{optproblem_single}
    \bm{\Phi}^*(k) = \argmax_{\substack{\bm{\Phi}(k)}} \Gamma_k(\bm{\Phi}(k)), \quad  \text{s.t.}\; \eqref{const_phaseshifters}.
\end{equation}
By defining $\bm{v}_k = \bm{w}_{{\rm U}_k}^T \bm{G}$ and $\bm{u}_k = \bm{H} \bm{w}_{{\rm g}_k}$ and re-writing the received signal power as 
\begin{equation}
    \left|\bm{v}_k\bm{\Phi}(k) \bm{u}_k\right|^2 = \left|\sum_{n = 1}^{N_{\rm I}} |[\bm{v}_k]_{n}||[\bm{u}_k]_{n}|e^{j(\angle[\bm{v}_k]_{n}+\theta_n+\angle[\bm{u}_k]_{n})}  \right|^2
\end{equation}
Then, it is sufficient to observe that the \ac{snr} is maximized when the phase shifts introduced by the \ac{irs} align the phase-shifts accumulated along the various paths, i.e., when 
\begin{equation}
    \theta_n = -(\angle[\bm{v}_k]_{n} + \angle[\bm{u}_k]_{n}), \quad \forall n.
\end{equation}

To overcome this interdependence between optimal \ac{irs} configurations and beamforming vectors we resort to an iterative alternate optimization approach. In particular, we first estimate the optimal beamforming vectors for a given \ac{irs} configuration using \eqref{eq:svd}. Then, we plug the derived beamformers into \eqref{optproblem_single} and obtain the corresponding optimal \ac{irs} configuration. We repeat this two-step procedure until convergence, which is usually reached in very few iterations.
For practical purposes, we assume the convergence is reached when the individual rate achieved in two consecutive iterations differs by less than $10^-4$.
It must be noted that the number of iterations needed grows with the number of considered antennas and \ac{irs} phase shifters. However, with our assumptions, convergence is always reached in less than $10$ iterations.
 
\subsection{Clustering-based TDMA scheduling}
\label{sec:clust_tdma}

For an approximated, but close-to-optimal solution to \eqref{optproblem}, we resort to a clustering-based approach.
Our proposed clustering algorithms estimate both the subsets of \acp{ue} $\mathcal{U}_1,\ldots,\mathcal{U}_{Z}$ and the set of respective \ac{irs} configurations $\mathcal{C}$.
In particular, they operate on the \textit{phase vector space}. Points to be clustered are identified by the \ac{irs} phase shifts vector, i.e.,
\begin{equation}
    \left[\angle{\left[\phi\right]_{0}},\ldots,\angle{\left[\phi\right]_{N_{\rm I}}}\right]^T = \left[\theta_{0},\ldots,\theta_{N_{\rm I}}\right]^T,
\end{equation}
which maps each \ac{irs} configuration $\bm{\Phi}$ to a point in a discrete grid on the continuous space $\left[0,2\pi\right]^{N_{\rm I}}$.

The general clustering-based procedure works as follows:
\begin{enumerate}[label=\arabic*.]
    \item find $\bm{\Phi}^*(k),\, \forall k$, i.e., the optimal \ac{irs} configurations for each \ac{ue}, 
    \item build the \ac{ue} subsets $\mathcal{U}_z, \, z = 1, \ldots, Z$ by using an arbitrary clustering algorithm,
    \item assign $\bm{\Phi}^{(z)}$ to all $k\in\mathcal{U}_z$
\end{enumerate}

\subsection{Capacity-Weighted Clustering}
\label{sec:clust_cwc}
With the aim of maximizing the average system sum rate, we now present the \ac{cwc} algorithm.
The algorithm performs a variable number of
iterations until reaching convergence, i.e., until the sum rate difference with the previous step is negligible.
Let $\bm{\Phi}_i^{(z)}$ be the centroid of the $z$-th cluster at iteration $i$.
\Acp{ue} are initially sorted in decreasing order of achievable \ac{snr}. The algorithm  selects the $Z$ \acp{ue} providing the best performance with their optimal \ac{irs} configurations and sets $\bm{\Phi}_1^{(z)} = \bm{\Phi}^{*}(k)$, $k = 1,\ldots, Z$, $z = 1,\ldots, Z$ as the initial centroids.
Then, each \ac{ue} $k > Z$ is assigned to $z^*_{k}$, defined as the cluster whose centroid provides the lowest rate difference with respect to the ideal configuration as
\begin{equation}
    z^* = \argmin_{\substack{z}}
    \log_2 \left( \frac{ 1+ \Gamma_k(\bm{\Phi}^*(k)) }
    {1{+} \Gamma_k(\bm{\Phi}_i^{(z)})} \right),
\end{equation}
where we exploit the well-known logarithm sum property.
After the assignments of all the remaining \acp{ue}, the coordinates of the centroids need to be updated.
At iteration $i+1$ the new centroid of cluster $z$ is computed as the average of all data points belonging to it, weighted by their achievable rate when adopting the centroid as 
\begin{equation}
    \bm{\Phi}_{i+1}^{(z)} = \frac{\sum_{k \in \mathcal{U}_z}  \bm{\Phi}^{*}(k) \log_2\left(1+\Gamma_k(\bm{\Phi}_{i}^{(z)})\right)}{\sum_{k \in \mathcal{U}_z}\log_2\left(1+\Gamma_k(\bm{\Phi}_{i}^{(z)})\right)}.
\end{equation}
We repeat this two-step procedure until convergence, which is usually reached in very few iterations, provided that the 
considered the number of antennas and \ac{irs} phase shifters is relatively limited.
However, in the case of massive \ac{mimo} systems, this procedure could reach high degrees of complexity, thus we also propose another low-complexity clustering algorithm, denoted as \ac{oscwc}. 

\subsection{One-Shot Capacity-Weighted Clustering}
As a low-complexity heuristic clustering solution to problem \eqref{optproblem} we propose \ac{oscwc}. As for \ac{cwc}, \acp{ue} are  sorted in decreasing order of achievable rate and the $Z$ \ac{irs} configurations making the \acp{ue} achieving highest rate are chosen as clusters centroids.
Then, instead of associating points to clusters and recomputing the coordinates of the centroids at each iteration, the algorithm stops 
upon the initial association.
Therefore, the  computed centroids are exactly the optimal configurations of the $Z$ \acp{ue} achieving the highest individual rate. 

\section{Numerical Results}\label{sec:numerical_results}

In this section, we assess via simulation the performance of an \ac{irs}-aided system with practical constraints.  
In the 2-D plane, we consider a \ac{umi} cell \cite{3gpp.38.901}, with the \ac{gnb} placed at the origin.
The coverage area is delimited by the 120° \ac{fov} of the \ac{gnb} and the cell radius, according to the specification, is fixed to $167$~m.
$K=100$ \acp{ue} are uniformly deployed \acp{ue} within the cell coverage area, to be served in the downlink by the \ac{gnb} assisted by an\ac{irs} with $(x, y)$ coordinates of $(75, 100)$~m. 
We remark that the direct links between \ac{gnb} and all \acp{ue} are assumed to exhibit severe attenuation due to blockage, thus being not exploitable for data transmission.
The \ac{gnb} is equipped with an $8\times8$ \ac{upa} antenna panel ($N_{\rm g}=64$) and all \acp{ue} with \acp{ula} of $N_{\rm U} = 2$ antennas. For the \ac{irs} size, if not specified in the following, we adopt a $40\times80$ reflective panel ($N_{\rm I}= 3200$).
The transmission power at the \ac{gnb} is set to $33$~dBm, while the noise power spectral density at the receivers is $-174$~dBm/Hz. Finally, the total system bandwidth is $100$~MHz.
We consider the \ac{3gpp} TR~38.901 spatial channel model~\cite{3gpp.38.901}, wherein channel matrices are computed based on the superposition of different clusters of rays, each one arriving (departing) to (from) the antenna arrays with specific angles and powers.
Finally, we assume that the \ac{gnb}-\ac{irs} link exhibits a \ac{los} path, while the \ac{irs}-\ac{ue} links are in \ac{nlos} conditions.

\subsection{Clustering Algorithms Benchmark}
The performance of our proposed scheduling strategies, i.e., \ac{cwc} and \ac{oscwc}, is compared to that achieved by the following traditional clustering algorithms.
\paragraph*{\ac{km}}
\ac{km} clustering~\cite{rokach2005clustering} aims at finding $Z$ disjoint clusters minimizing the within-cluster sum of squares. To perform \ac{km} clustering, we here consider the well-known Lloyd algorithm \cite{Kmeans}, which randomly selects $Z$ points in the space of phase vectors as the initial centroids. Then, it assigns each data point to the closest centroid, i.e., the one with the smallest squared Euclidean distance. The set of centroids is then re-computed as the average of all data points that belong to each cluster. These steps are repeated until either convergence is met, or a maximum number of iterations (here fixed to $100$) is reached.
\paragraph*{Agglomerative \ac{hc}}
The agglomerative \ac{hc}~\cite{murtagh2012algorithms} is a clustering technique that partitions a set of data points into disjoint clusters by iteratively merging points into clusters, until the target number of partitions is met.
In particular, the clusters are initialized as the optimal phase vectors, which thus act as the respective centroids. Then, the average Euclidean distance between all pairs of data points in any two clusters is evaluated. The closest pair of clusters is then merged into a new single cluster whose centroid is computed as the average of all its data points. The procedure is repeated until the number of clusters is equal to $Z$. 


Finally, we also consider two bounds on the scheduling performance.
The former is an achievable lower bound based on \textit{random clustering}, where the \acp{ue} are randomly partitioned into $Z$ clusters, and each cluster centroid is computed as the average of all data points in the cluster. 
Thus, it is reasonable to expect that any sensible algorithm performs better than this trivial solution.
The latter, denoted as \textit{unclustered scheduling}, assumes that all \acp{ue} are served with their optimal \ac{irs} configuration. This is clearly an upper bound that violates the constraint on the minimum reconfiguration period, but can be regarded as the limit case when $Z = K$, thus all \acp{ue} belong to a cluster with cardinality one.

\subsection{Scheduling Performance}


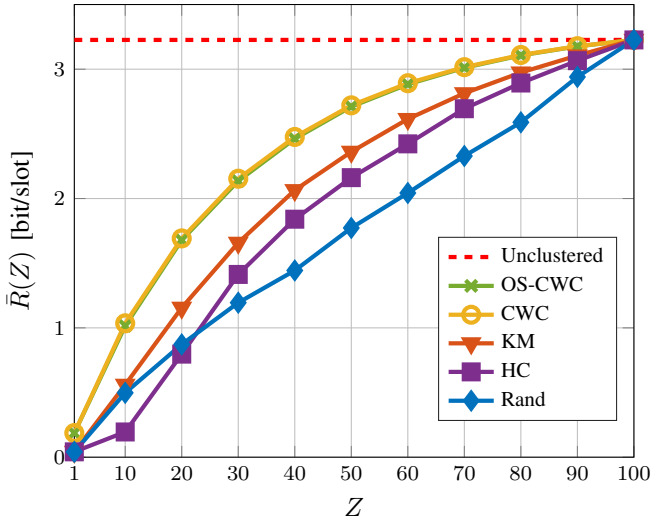
\begin{figure}
    \centering
    \setlength\fwidth{0.84\columnwidth}
    \setlength\fheight{0.68\columnwidth}
%
%
\definecolor{mycolor1}{rgb}{0.46600,0.67400,0.18800}%
\definecolor{mycolor2}{rgb}{0.92941,0.69412,0.12549}%
\definecolor{mycolor3}{rgb}{0.85000,0.32500,0.09800}%
\definecolor{mycolor4}{rgb}{0.49400,0.18400,0.55600}%
\definecolor{mycolor5}{rgb}{0.00000,0.44700,0.74100}%

\pgfplotsset{every tick label/.append style={font=\footnotesize}}

\begin{tikzpicture}

\begin{axis}[%
    width=\fwidth,
    height=\fheight,
    at={(0\fwidth,0\fheight)},
    scale only axis,
    ylabel style={font=\normalsize},
    xlabel style={font=\normalsize},
    xmin=1,
    xmax=100,
    xtick={  1,  10,  20,  30,  40,  50,  60,  70,  80,  90, 100},
    xlabel={$Z$},
    ymin=0,
    ymax=3.5,
    ylabel={$\bar{R}(Z)$ [bit/slot]},
    axis background/.style={fill=white},
    xmajorgrids,
    ymajorgrids,
    legend style={at={(0.65,0.08)}, anchor=south west, legend cell align=left, align=left, font=\footnotesize, draw=white!15!black}
]
\addplot [color=red, dashed, ultra thick]
  table[row sep=crcr]{%
1	3.22636030711816\\
10	3.22636030711816\\
20	3.22636030711816\\
30	3.22636030711816\\
40	3.22636030711816\\
50	3.22636030711816\\
60	3.22636030711816\\
70	3.22636030711816\\
80	3.22636030711816\\
90	3.22636030711816\\
100	3.22636030711816\\
};
\addlegendentry{Unclustered}

\addplot [color=mycolor1, ultra thick, mark size=3.0pt, mark=x, mark options={solid, mycolor1}]
  table[row sep=crcr]{%
1	0.186961533645684\\
10	1.0152592560466\\
20	1.67843858949881\\
30	2.13578290193494\\
40	2.46200653803852\\
50	2.71045619061224\\
60	2.883134526476\\
70	3.01086866039245\\
80	3.10672443982847\\
90	3.17648068354771\\
100	3.22636033555578\\
};
\addlegendentry{OS-CWC}

\addplot [color=mycolor2, ultra thick, mark size=3.0pt, mark=o, mark options={solid, mycolor2}]
  table[row sep=crcr]{%
1	0.186961533645684\\
10	1.03521428168236\\
20	1.69176147955796\\
30	2.15320699147614\\
40	2.47608621726546\\
50	2.72047928421574\\
60	2.89170951854995\\
70	3.01575869535762\\
80	3.1099125020428\\
90	3.17792138331619\\
100	3.22636033555578\\
};
\addlegendentry{CWC}

\addplot [color=mycolor3, ultra thick, mark size=3.0pt, mark=triangle*, mark options={solid, rotate=180, mycolor3}]
  table[row sep=crcr]{%
1	0.0415349763175782\\
10	0.559790037597512\\
20	1.15554172928068\\
30	1.65776752858959\\
40	2.06463443771061\\
50	2.36136991998179\\
60	2.61465958166523\\
70	2.81393100222055\\
80	2.97419800993532\\
90	3.10580205927758\\
100	3.22636033555578\\
};
\addlegendentry{KM}

\addplot [color=mycolor4, ultra thick, mark size=2.8pt, mark=square*, mark options={solid, mycolor4}]
  table[row sep=crcr]{%
1	0.0415349763175782\\
10	0.195105020502851\\
20	0.796944741483586\\
30	1.41312202841851\\
40	1.84018666273201\\
50	2.1615541490659\\
60	2.42277874328523\\
70	2.69449364122376\\
80	2.89319121193064\\
90	3.06588537511251\\
100	3.22636033555578\\
};
\addlegendentry{HC}

\addplot [color=mycolor5, ultra thick, mark size=3.0pt, mark=diamond*, mark options={solid, mycolor5}]
  table[row sep=crcr]{%
1	0.0415349763175782\\
10	0.497815823440019\\
20	0.872268754390124\\
30	1.19464046333939\\
40	1.4436398384112\\
50	1.77258904552913\\
60	2.04268297195839\\
70	2.32882877840938\\
80	2.58913984738289\\
90	2.94011811054444\\
100	3.22636033555578\\
};
\addlegendentry{Rand}

\end{axis}
\end{tikzpicture}%
    \caption{Average sum-rate as a function of the maximum number of clusters allowed $Z$.}
    \label{fig:single_carrier_rate}
\end{figure}

Fig.~\ref{fig:single_carrier_rate} shows the average rate per slot as a function of the number of clusters $Z$. It can be observed that all the scheduling policies are increasing with respect to $Z$, and they converge to the unclustered policy. This is motivated by the fact that a higher number of clusters leads to a smaller intra-cluster average distance, which eventually becomes $0$ for all clusters when $Z=K$. In turn, this metric relates to ``how far" a given \ac{ue} ideal configuration is from the centroid, i.e., the configuration which will be used by all the \acp{ue} which belong to the cluster. 
Among the considered clustering policies, \ac{cwc} and \ac{oscwc}  provide the highest sum-rate thanks to the bias towards \acp{ue} which achieve a good signal quality when served using their optimal \ac{irs} configuration. 
Finally, the gap between \ac{cwc} and \ac{oscwc} is negligible, which suggests that \acp{ue}' ideal configurations belong to well-defined isolated regions of the phase-vector space. As a consequence, a single iteration in the clustering procedure is enough to achieve good performance, which implies a good scalability of the proposed technique.

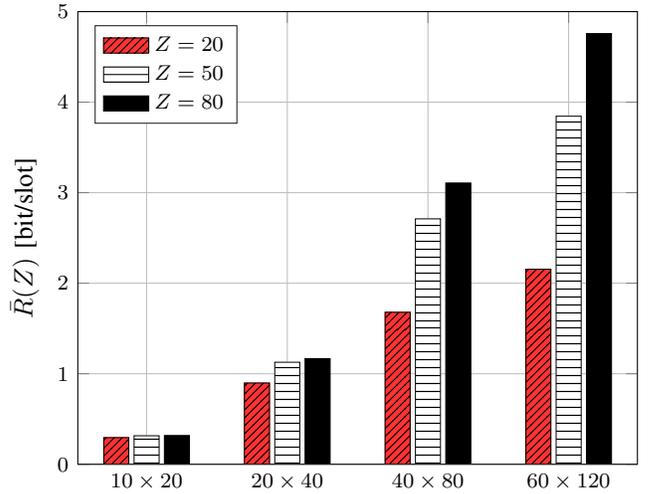
\begin{figure}
    \centering
    \setlength\fwidth{0.84\columnwidth}
    \setlength\fheight{0.68\columnwidth}
%
%

\definecolor{mycolor1}{rgb}{1,0.2,0.2}%
\definecolor{mycolor2}{rgb}{1,1,1}%
\definecolor{mycolor3}{rgb}{0,0,0}%

\pgfplotsset{every tick label/.append style={font=\footnotesize}}
\begin{tikzpicture}

\begin{axis}[%
    width=\fwidth,
    height=\fheight,
    at={(0\fwidth,0\fheight)},
scale only axis,
bar shift auto,
xmin=0.511111111111111,
xmax=4.48888888888889,
xtick={1,2,3,4},
xticklabels={{$10\times20$},{$20\times40$},{$40\times80$},{$60\times120$}},
ymin=0,
ymax=5,
ylabel style={font=\normalsize},
xlabel style={font=\normalsize},
ylabel={$\bar{R}(Z)$ [bit/slot]},
axis background/.style={fill=white},
xmajorgrids,
ymajorgrids,
legend style={at={(0.03,0.97)}, anchor=north west, legend cell align=left, font=\footnotesize, align=left, draw=white!15!black}
]
\addplot[ybar, bar width=0.178, preaction={fill, mycolor1}, very thin, pattern={north east lines}, pattern color=black, draw=black, area legend] table[row sep=crcr] {%
1	0.294282273682828\\
2	0.897442120724023\\
3	1.67898173789839\\
4	2.15457536084892\\
};
\addplot[forget plot, color=white!15!black] table[row sep=crcr] {%
0.511111111111111	0\\
4.48888888888889	0\\
};
\addlegendentry{$Z=20$}

\addplot[ybar, bar width=0.178, preaction={fill, mycolor2}, very thin, pattern={horizontal lines}, pattern color=black, draw=black, area legend] table[row sep=crcr] {%
1	0.315125531676461\\
2	1.12720090444033\\
3	2.71088877612277\\
4	3.84465836645518\\
};
\addplot[forget plot, color=white!15!black] table[row sep=crcr] {%
0.511111111111111	0\\
4.48888888888889	0\\
};
\addlegendentry{$Z=50$}

\addplot[ybar, bar width=0.178, fill=mycolor3, draw=black, area legend] table[row sep=crcr] {%
1	0.317547194929616\\
2	1.16709740636668\\
3	3.10666899658725\\
4	4.75673149618294\\
};
\addplot[forget plot, color=white!15!black] table[row sep=crcr] {%
0.511111111111111	0\\
4.48888888888889	0\\
};
\addlegendentry{$Z=80$}
\end{axis}

\end{tikzpicture}%
    \caption{Average sum-rate as a function of the number of \ac{irs} radiating elements.}
    \label{fig:single_carrier_vs_size}
\end{figure}

Fig.\ \ref{fig:single_carrier_vs_size} depicts the impact of the number of \ac{irs} radiating elements on the system performance, when considering the \ac{cwc} policy. As expected, the achievable rate increases as we use bigger \acp{irs}, regardless of the number of clusters. 
Furthermore, it can be noted that varying the number of reflecting elements has an impact on the number of clusters that are needed to provide the maximum achievable rate as well. Indeed, the number of possible \ac{irs} configurations increases as we consider arrays featuring additional antenna elements, as the reflected beams get progressively narrower. In turn, this decreases the likelihood of \acp{ue} exhibiting the same (or similar) ideal configurations. We remark that our proposed solution is able to almost provide the optimal rate with as few as $20$ clusters for small-sized \acp{irs}, i.e., featuring $10\times20$ or $20\times40$ arrays.

\begin{figure}
    \centering
    \setlength\fwidth{0.84\columnwidth}
    \setlength\fheight{0.68\columnwidth}
%
%
\definecolor{mycolor1}{rgb}{0.92900,0.69400,0.12500}%

\pgfplotsset{every tick label/.append style={font=\footnotesize}}
\begin{tikzpicture}

\begin{axis}[%
    width=\fwidth,
    height=\fheight,
    at={(0\fwidth,0\fheight)},
    scale only axis,
    ylabel style={font=\normalsize},
    xlabel style={font=\normalsize},
    xmin=1,
    xmax=100,
    xtick={1,  10,  20,  30,  40,  50,  60,  70,  80,  90, 100},
    xlabel={$Z$},
    ymin=0,
    ymax=3.5,
    ylabel={$\bar{R}(Z)$ [bit/slot]},
    axis background/.style={fill=white},
    xmajorgrids,
    ymajorgrids,
    legend style={at={(0.97,0.03)}, anchor=south east, legend cell align=left, align=left, font=\footnotesize, draw=white!15!black}
]
\addplot [color=mycolor1, ultra thick, mark size=3.3pt, mark=o, mark options={solid, mycolor1}]
  table[row sep=crcr]{%
1	0.186961533645684\\
10	1.0152592560466\\
20	1.67843858949881\\
30	2.13578290193494\\
40	2.46200653803852\\
50	2.71045619061224\\
60	2.883134526476\\
70	3.01086866039245\\
80	3.10672443982847\\
90	3.17648068354771\\
100	3.22636033555578\\
};
\addlegendentry{Unquantized}

\addplot [color=red, dashed, ultra thick, mark size=3.0pt, mark=asterisk, mark options={solid, red}]
  table[row sep=crcr]{%
1	0.178594071401424\\
10	0.963043452655795\\
20	1.57355538829977\\
30	1.9806446516572\\
40	2.25959184427751\\
50	2.47001915082096\\
60	2.60989348088454\\
70	2.70991425483463\\
80	2.78256858578436\\
90	2.83499442890487\\
100	2.87139085984396\\
};
\addlegendentry{$b =5$}

\addplot [color=black, dashdotted, ultra thick, mark size=3.0pt,  mark=triangle, mark options={solid, rotate=0, black}]
  table[row sep=crcr]{%
1	0.165642712448702\\
10	0.876093115978339\\
20	1.3963630797225\\
30	1.72573931366248\\
40	1.93655594832429\\
50	2.08813319833202\\
60	2.18314021038932\\
70	2.24718093361842\\
80	2.29129837185638\\
90	2.3226269849661\\
100	2.34394490011499\\
};
\addlegendentry{$b =2$}

\addplot [color=blue, dotted, ultra thick, mark size=2.8pt, mark=otimes, mark options={solid, blue}]
  table[row sep=crcr]{%
1	0.132794636294557\\
10	0.666995147336268\\
20	0.986164138458976\\
30	1.14968788219159\\
40	1.23400200306336\\
50	1.28705198557476\\
60	1.31694978453894\\
70	1.3345910604975\\
80	1.34573227755667\\
90	1.35327478268977\\
100	1.35801889694612\\
};
\addlegendentry{$b =1$}
\end{axis}
\end{tikzpicture}%
    \caption{Average sum-rate as a function of the \ac{irs} phase-shifters quantization resolution.}
    \label{fig:single_carrier_vs_bits}
\end{figure}
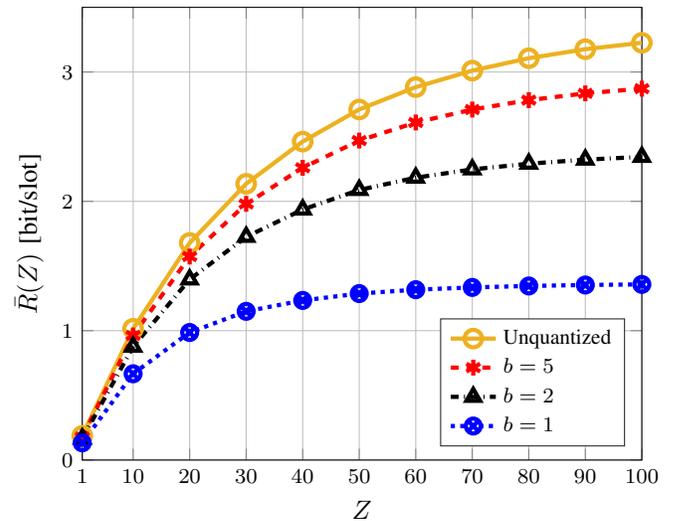

Finally, Fig.\ref{fig:single_carrier_vs_bits} shows the average sum-rate as a function of the maximum number of clusters computed by the \ac{cwc} algorithm, for different numbers of bits $b$ used for the \ac{irs} phase shifts quantization. Results show that considering non-ideal phase-shifters leads to a considerable rate degradation when \acp{ue} are in this particular \ac{snr} region. In particular, considering a $b=1$ quantization at the \ac{irs} incurs up to a $60\%$ reduction in the achieved rate with respect to the continuous case. Instead, such a gap is not as dramatic when considering higher resolution phase-shifters. 

\section{Conclusions}\label{sec:conclusions}
In this paper, we have considered a \ac{mimo} communication system, in which a \ac{gnb} that serves multiple \acp{ue} experiences a deep blockage and is thus assisted by an \ac{irs} possibly having practical constraints on his configuration period. We have considered a TDMA scheduling of downlink transmissions, and we have formulated an optimization problem that aims at maximizing the average system rate subject to a fixed number of \ac{irs} reconfigurations per radio time frame.
We have mitigated the performance degradation caused by such a limitation by proposing clustering-based scheduling policies, which group \acp{ue} with similar ideal IRS configurations. This allows to reduce the number of configurations as all the \acp{ue} belonging to a specific cluster are served with the same IRS parameters.

We have analyzed the sum-rate performance of the proposed clustering-based approaches and highlighted the benefit of adopting a rate-driven clustering with respect to the traditional \ac{km} or \ac{hc} algorithms. The obtained results show that our approach is effective in guaranteeing up to $85$\% of the throughput obtained by an ideal deployment (with no reconfiguration constraints) while providing a $50$\% reduction in the number of \ac{irs} reconfigurations.

\balance

\bibliographystyle{IEEEtran}
\bibliography{IEEEabrv,biblio}

\end{document}